\documentclass[12pt]{article}
\usepackage{amsfonts}

\begin{document}

\title{Effective Hamiltonian with position dependent mass and ordering problem}

\author{V. M. Tkachuk$^1$, O. Voznyak\\
Department for Theoretical Physics,\\ Ivan Franko National
University of Lviv,\\ 12 Drahomanov St., Lviv, 79005, Ukraine\\
$^1$e-mail:
voltkachuk@gmail.com}

\maketitle

\begin{abstract}
We derive the effective low energy Hamiltonian for the tight-binding model with the hopping integral slowly varying along the chain.
The effective Hamiltonian contains the kinetic energy with position dependent mass, which is inverse to the hopping integral, and effective potential energy.
Changing of ordering in the kinetic energy  leads to change of the effective potential energy and leaves the Hamiltonian the same one.
Therefore, we can choose arbitrary von Roos ordering parameters in the kinetic energy without changing the Hamiltonian.
Moreover, we propose a more general form for the kinetic energy than that of von Roos, which nevertheless together with the effective potential energy represent the same Hamiltonian.

Key words: tight-binding model, effective Hamiltonian, position dependent mass, ordering problem
\end{abstract}

\section{Introduction}
Particles with position-dependent mass in
quantum theory have attracted attention for the last few
decades due to their applications in condensed-
matter physics.
The studies of quantum mechanical systems with position-dependent mass raise an
important conceptual problem, such as the ordering ambiguity of the momentum and
mass operators in the kinetic energy term.
In the last few
decades a little progress in solving the ordering problem has been made.
Few methods for solving this problem have been proposed.
In \cite{Ein190,Ein290} the ordering problem was studied by comparison of exact results with effective-mass results for solvable test models.
In this relation we would like also to cite paper \cite{Sou00} where the authors discussed the problem of solvability and
ordering ambiguity in quantum mechanics.
In our papers \cite{Que04,Bag05} we generalized exactly solvable shape-invariant potentials for the case of position dependent mass.
It is interesting to note that the Coulomb problem is exactly solved in the case of arbitrary ordering of momentum and mass in the kinetic energy and some special dependence of mass on the coordinates \cite{Bag05}.
In \cite{Hab13} it was shown that the classical and quantum mechanical correspondence may play a basic role in the fixation of the ordering-ambiguity parameters. The authors of
paper \cite{Mus07} revised the ordering ambiguity via position
dependent mass pseudo-momentum operators.
A physical method for solving the ordering problem was proposed in \cite{You89}.
In that paper, the effective low energy
Hamiltonian for a crystal with a slowly varying
inhomogeneity was extracted. The author made the calculation accurate to the
second order in the gradient operator. Finishing this short review on ordering problem, we would like to
cite recent paper \cite{Vub14} on this subject (see also references therein).

In this paper we use a physical method for obtaining an effective Hamiltonian, namely, we start from the microscopic tight-binding model with the hopping integral varying along the chain. Then, considering slowly varying
hopping integral, we get effective low energy
Hamiltonian.
In contrast to the method proposed in \cite{You89} we consider site representation for eigenvalue equation and do not use the Bloch basis.
Therefore, our method is more direct and simple than the method proposed in \cite{You89}.

\section{Effective Hamiltonian for the inhomogeneous tight binding model}
We consider nonuniform one particle tight-binding model with Hamiltonian
\begin{eqnarray}\label{TBM}
H=\sum_i J_i(|i+1\rangle\langle i|+|i\rangle\langle i+1|)  +\sum_i \epsilon_i |i\rangle\langle i|,
\end{eqnarray}
where $J_i$ and $\epsilon_i$ are the hopping integral and the energy on the site $i$, respectively.
The state vectors $|i\rangle$ corresponding to different sites $i$ are orthogonal.

Let us consider the stationary Schr\"odinger equation
\begin{eqnarray}
H|\psi\rangle=E|\psi\rangle,
\end{eqnarray}
solution of which can be found in the form
\begin{eqnarray}
|\psi\rangle=\sum_i c_i|i\rangle.
\end{eqnarray}
Equation for $c_i$ reads
\begin{eqnarray} \label{Eqc}
J_ic_{i+1}+ J_{i-1}c_{i-1}+\epsilon_i c_i=Ec_i.
\end{eqnarray}
The dependence of $c_i$ on the position of site $i$ can be written as follows
\begin{eqnarray}
c_i=\psi(x_i),
\end{eqnarray}
where $x_i$ is the position of site $i$ and $a$ is the distance between two neighbor sites.
We suppose that $\psi$ is varying very slowly at distances of order of $a$.
This corresponds to the low energy levels.
Then in the second order over $a$ we can write
\begin{eqnarray}
c_{i+1}=\psi(x_i+a)=\psi(x_i)+\psi'(x_i)a+{1\over 2}\psi''(x_i)a^2,\\
c_{i-1}=\psi(x_i-a)=\psi(x_i)-\psi'(x_i)a+{1\over 2}\psi''(x_i)a^2,\\
J_i=J(x_i+a)=J(x_i)+J'(x_i)a+{1\over 2}J''(x_i)a^2,
\end{eqnarray}
where $\psi'(x)=d\psi(x)/dx$, $\psi''(x)=d^2\psi(x)/dx^2$.
Substituting these expansions into (\ref{Eqc}), we find
\begin{eqnarray}\nonumber
a^2\left(J(x)\psi''(x)+ J'(x)\psi'(x)+ {1\over2} J''(x)\psi(x)\right)-
aJ'(x)\psi(x) \\
+2J(x)\psi(x)+\epsilon(x)\psi(x)=E\psi(x),
\end{eqnarray}
here instead of $x_i$ we write $x$ which can be treated as a continuous variable.
This equation can be considered as the Schrodinger equation with Hamiltonian
\begin{eqnarray}\label{Hx}
H=a^2\left(J(x){d^2\over dx^2}+J'(x){d\over dx}\right) +{1\over 2}a^2J''(x)-a J'(x)+2J(x)+\epsilon(x).
\end{eqnarray}
One can verify that
\begin{eqnarray}
\left(J(x){d^2\over dx^2}+J'(x){d\over dx}\right)={d\over dx}J(x){d\over dx},
\end{eqnarray}
and thus Hamiltonian (\ref{Hx}) can be written explicitly in  hermitian form
\begin{eqnarray}\label{HxHerm}
H=a^2{d\over dx}J(x){d\over dx} +{a^2\over 2}J''(x)-a J'(x)+2J(x)+\epsilon(x).
\end{eqnarray}
This is the effective low energy
Hamiltonian corresponding to tight binding model (\ref{TBM}) with slowly varying
hopping integral which is inverse to the mass. Note that the ordering in the kinetic energy obtained here is the same as was derived in
\cite{You89}.

\subsection{Different types of ordering}
One can think that the first term in (\ref{HxHerm}) can be treated as kinetic energy and the second term as potential energy.
But in fact we can rewrite the same Hamiltonian (\ref{HxHerm}) in other form (see, for instance, \cite{Que04,Bag05}).
Using identity
\begin{eqnarray}
{1\over 2}\left(J^{\alpha}(x){d\over dx}J^{\beta}(x){d\over dx}J^{\gamma}(x)+J^{\gamma}(x){d\over dx}J^{\beta}(x){d\over dx}J^{\alpha}(x)\right)=\\
J(x){d^2\over dx^2}+J'(x){d\over dx}+{1\over 2}(\alpha +\gamma)J''(x)-\alpha\gamma{(J'(x))^2\over J(x)},
\end{eqnarray}
where $\alpha +\beta+\gamma=1$,
we can write Hamiltonian (\ref{Hx}) as a sum of kinetic and potential energy
\begin{eqnarray}\label{H}
H=T+U.
\end{eqnarray}
Here operators of kinetic and potential energies read
\begin{eqnarray}\label{T}
T={a^2\over 2}\left(J^{\alpha}(x){d\over dx}J^{\beta}(x){d\over dx}J^{\gamma}(x)+J^{\gamma}(x){d\over dx}J^{\beta}(x){d\over dx}J^{\alpha}(x)\right),\\ \label{U}
U={a^2\over 2}(1-\alpha-\gamma)J''(x)+a^2\alpha\gamma{(J'(x))^2\over J(x)}-aJ'(x)+2J(x)+\epsilon(x),
\end{eqnarray}
respectively. Note that this form of kinetic energy was suggested by von Roos \cite{Roos83}.

So, the same Hamiltonian (\ref{HxHerm}) can be written in form (\ref{H}) with different ordering in the kinetic energy.
According to (\ref{U}) changing of the ordering in (\ref{T}) leads to changing of the potential energy in such a way that sum of the kinetic energy
and potential one does not depend on the ordering. We can choose different orderings depending on the convenience and obtain the same Hamiltonian written in different forms. For instance, for $\beta=1$, $\alpha=\gamma=0$ we obtain the Hamiltonian in form (\ref{HxHerm}),
for $\alpha=1$, $\gamma=\beta=0$ Hamiltonian (\ref{HxHerm}) reads
\begin{eqnarray}\label{H1}
H={a^2\over2}\left(J(x){d^2\over dx^2}+{d^2\over dx}J(x)\right) -a J'(x)+2J(x)+\epsilon(x).
\end{eqnarray}

Note that von Roos form (\ref{T}) is not the most general form of the kinetic energy with position dependent mass. We propose more general form as follows
\begin{eqnarray}\label{TG}
T_G={a^2\over 2}\left(J_1(x){d\over dx}J_2(x){d\over dx}J_3(x)+J_3(x){d\over dx}J_2(x){d\over dx}J_1(x)\right),
\end{eqnarray}
where three functions $J_1(x)$, $J_2(x)$, $J_3(x)$ satisfy the condition $J_1(x)J_2(x)J_3(x)=J(x)$.
One can verify that (\ref{TG}) is equal to
\begin{eqnarray}
T_G=a^2\left(J(x){d^2\over dx^2}+J'(x){d\over dx}\right)+\\ \nonumber
+{a^2\over 2}\left[J_1(x)(J_2(x)J'_3(x))'+J_3(x)(J_2(x)J'_1(x))'\right].
\end{eqnarray}
It is important to note that the first term in the right-hand side of this expression is the same as the kinetic term in
effective Hamiltonian (\ref{Hx}). This allows us to write effective Hamiltonian (\ref{Hx}) as a sum of kinetic energy (\ref{TG})
and potential energy
\begin{eqnarray} \label{UG}
U={1\over 2}a^2J''(x)-a J'(x)+2J(x)+\epsilon(x)-\\ \nonumber
-{a^2\over 2}\left[J_1(x)(J_2(x)J'_3(x))'+J_3(x)(J_2(x)J'_1(x))'\right].
\end{eqnarray}

\section{Conclusions}

In this paper we have considered the  tight-binding model with the hopping integral slowly varying along the chain.
We have obtained the effective low energy Hamiltonian (\ref{HxHerm}) which contains the kinetic energy with position dependent mass
and the effective potential energy.
We have concluded that the changing of the ordering in the kinetic energy leads to change of the effective potential energy and leaves Hamiltonian the same.
So, the same Hamiltonian can be written in general form (\ref{H}) as a sum of kinetic energy with arbitrary ordering of mass and momentum (\ref{T}) and effective potential energy (\ref{U}).
We have also proposed more general form for kinetic energy (\ref{TG}) with position dependent mass than von Roos one which nevertheless together with effective potential energy (\ref{UG}) represent the same Hamiltonian.
Therefore, there is no sense to consider the ordering problem in the kinetic energy without taking into account the potential energy at least for the model studied in this paper.
We can ask only the question what is the effective Hamiltonian, namely, the kinetic energy together with the potential one for a particle with position dependent mass.
In this paper, we have given the answer to this question for the tight-binding model with slowly varying hopping integral
and have obtained effective Hamiltonian (\ref{HxHerm}).

\end{document}